\documentclass{osa-article}

\journal{osajournal}

\articletype{Research Article}

\begin{document}

\title{Field-based Design of a Resonant Dielectric Antenna for Coherent Spin-Photon Interfaces }

\author{Linsen Li\authormark{1}, Hyeongrak Choi\authormark{1}, Mikkel Heuck\authormark{1}, and Dirk Englund\authormark{1}}

\address{\authormark{1}Research Laboratory of Electronics, Massachusetts Institute of Technology, Cambridge, MA 02139, USA}

\email{\authormark{*}mheuck@mit.edu} 
\email{\authormark{*}englund@mit.edu} 



\begin{abstract}
We propose a field-based design for dielectric antennas to interface diamond color centers with a Gaussian propagating far field. This antenna design enables an efficient spin-photon interface with a Purcell factor exceeding 400 and a 93\% mode overlap to a 0.4 numerical aperture far-field Gaussian mode. The antenna design is robust to fabrication imperfections, such as variations in the dimensions of the dielectric perturbations and the emitter dipole location. The field-based dielectric antenna design provides an efficient free-space interface to closely packed arrays of quantum memories for multiplexed quantum repeaters, arrayed quantum sensors, and modular quantum computers.
\end{abstract}

\section{Introduction}
Color centers in diamond, such as the nitrogen-vacancy (NV) centers, are promising solid-state qubit systems due to their long spin coherence time and stable optical transition~\cite{NV_Time}. Moreover, recently studied group IV-vacancy centers in diamond such as silicon-vacancy (SiV) centers~\cite{SiV}, germanium-vacancy (GeV) centers~\cite{GeV}, and tin-vacancy (SnV) centers~\cite{SnV,SnV_matt} have resilient optical transitions~\cite{DFT_GIV}. Quantum applications such as quantum network~\cite{Hanson18}, computing~\cite{choi19}, and sensing~\cite{cappellaro2016stable}, demand efficient coupling of quantum emitters to free-space propagating fields at cryogenic temperature. However, the free-space collection from the diamond color center is inefficient due to the large index mismatch between diamond and air. Recent photonic structures for improving free-space coupling include macroscopic solid immersion lenses produced by focused ion beam milling~\cite{hadden2010strongly, trojak2018combined}, recessed circular grating~\cite{GaAs_Nat_Nano,zheng2017chirped}, metal plasmonic grating~\cite{choy2013spontaneous, sensing}, metallic bow-tie antenna~\cite{M_antenna}, nanopatch antenna~\cite{bogdanov2020ultrafast}, metasurfaces~\cite{huang2019monolithic}, and diamond nanopillars~\cite{babinec2010bright}. However, none of these designs combines a high Purcell effect and directional emission to jointly optimize spectral and spatial collection.

Here, we introduce a design process for a dielectric antenna as a spin-photon interface to direct emission into the desired far-field mode. In particular, we introduce a transfer-matrix approach for 3D dielectric antenna design, combining gratings for directional mode coupling and cavity for Purcell enhancement. We apply this field-based design recipe to develop a diamond antenna that simultaneously achieves a Purcell factor of 420 and a 93\% mode overlap with a 0.4 numerical aperture (NA) Gaussian beam, which has a 99\% collection efficiency within an NA of 0.5. Thus, we estimate that the spin-photon interface efficiency can improve $\gtrsim 300$ times compared with an NV dipole embedded in a 150 nm thick diamond membrane without nanostructures. Dielectric antennas, unlike metallic antennas \cite{M_antenna}, do not suffer from ohmic loss and quenching. The surface charge and spin noise are also alleviated by placing the closest etched surface more than one wavelength away from the dipole emitter \cite{choi2019cascaded}. We believe this quantum emitter antenna structure to be of great utility for quantum applications.


\section{Antenna Design}

As a figure of merit, we consider a coherent spin-photon interface efficiency $ \eta= \eta_1 \eta_2 $, where $\eta_1$ denotes the spin-antenna interface efficiency and $\eta_2$ is the antenna efficiency. $\eta_1$ is defined as
\begin{equation}
\eta_1 = \frac{\eta_0\times F_p}{\eta_0\times F_p+1-\eta_0},
\end{equation}
where $\eta_0$ denotes the radiation efficiency into the zero-phonon line (ZPL). The values of $\eta_0$ for NVs, SiVs, and SnVs are 3\%, 7\%, and 32\%, respectively \cite{DFT_GIV}. $F_p$ is the Purcell Factor, which increases the spontaneous emission rate in the ZPL \cite{Purcell}. 

The antenna far field $\vec{E}_\text{far}=E_r\vec{r}+E_\theta\vec{\theta}+E_\phi\vec{\phi}$ is calculated on a hemispherical surface located $r_0=1$ m away from the center dipole source with finite-difference time-domain (FDTD) simulations using Lumerical. We expect that $E_r$ to be zero because the electric far field is perpendicular to the direction of propagation. Assuming a monochromatic antenna far field with angular frequency $\omega$ and corresponding free-space wavelength $\lambda$, $\vec{E}_\text{far}(r_0,\theta,\phi,t)=C_\text{far}\exp[-i(\omega t- \frac{2\pi r_0}{\lambda})]\vec{e}_\text{far}(\theta,\phi)$, where $C_\text{far}$ is the electric far-field amplitude. $\vec{e}_\text{far}(\theta,\phi) $ is the dimensionless normalized far field. $\vec{e}_\text{far}(\theta,\phi)$ satisfies $\int_0^{2\pi} \int_0^\pi |\vec{e}_\text{far}(\theta,\phi)|^2 \sin\theta \text{d}\theta \text{d}\phi =T_z=P_z/(2P_x+2P_y+P_z)$, where $T_z$ is the fraction of the emitted power propagating in the +$z$ direction. $P_x$, $P_y$, and $P_z$ are the radiated power propagating in the $x$, $y$, and $z$ direction, respectively (fig. 1(a)). The target far field  $\vec{E}_\text{tar}(r_0,\theta,\phi,t)$ has the same notation except the subscript.

The antenna efficiency is the square of the mode overlap between $\vec{e}_\text{far}(\theta,\phi)$ and $\vec{e}_\text{tar}(\theta,\phi)$, which we define as
\begin{equation}
\eta_2 =\left\lvert \int_0^{2\pi}\int_0^{\pi} \vec{e}_\text{far}(\theta,\phi) \cdot \vec{e}_\text{tar}^{~*}(\theta,\phi) \sin\theta \text{d}\theta \text{d}\phi \right\rvert ^2 .
\end{equation}
We use a polarized Gaussian beam with NA = 0.4 as the target far field, $\vec{e}_\text{tar}(\theta,\phi) = 2.10\times \exp(-\tan^2\theta/0.4^2)\vec{y}$, which has 99\% of its electromagnetic energy within an NA of 0.5 under paraxial approximation. Here, $\vec{e}_\text{tar}(\theta,\phi)$ satisfies $\int_0^{2\pi} \int_0^\pi |\vec{e}_\text{tar}(\theta,\phi)|^2 \sin\theta \text{d}\theta \text{d}\phi =1.$ We also consider the higher order correction for the polarized Gaussian beam \cite{levy2016weakly}, $\vec{e}_\text{tar}(\theta,\phi) = 2.14\times \exp(-\tan^2\theta/0.4^2)[(1-\tan^2\theta \sin^2\phi)\vec{y}-\tan^2\theta\sin\phi\cos\phi\vec{x}-\tan\theta\sin\phi(1-\tan^2\theta/2)\vec{z}]$. It has less than 1\% variation in our final efficiency calculation.

We summarize the field-based antenna design recipe as follows, providing the details of each step in the supplementary material:
\begin{figure}[htbp]
\centering
\includegraphics[width=13cm]{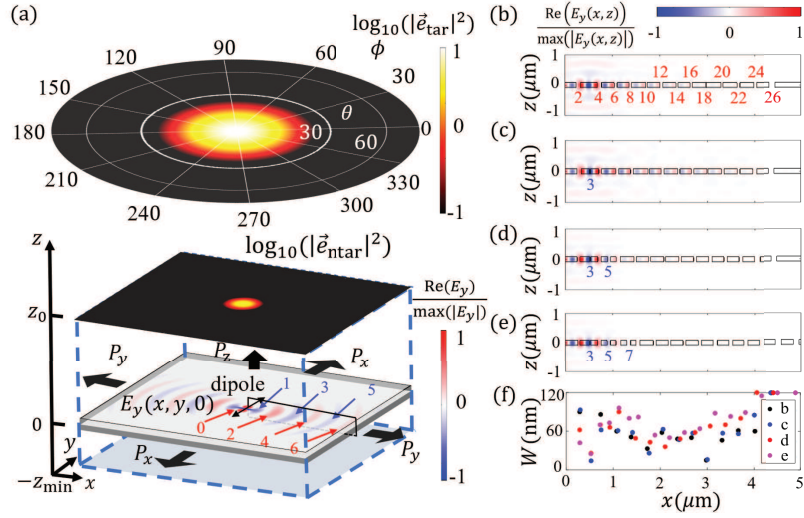}
\caption{Illustration of the field-based antenna design recipe. (a) 2D plots of the target far field $\log_{10}(|\vec{e}_{\text{tar}}|^2)$ (top), the target near field $\log_{10}(|\vec{e}_\text{ntar}|^2)$ (middle), and illustration of the unpatterned diamond slab with the dipole field overlaid with $\text{Re}(E_y)/\max(|E_y|)$ labeled with phase front number, assuming a perfect reflector at the $z = -z_\text{min} $ plane in the final structure. (b-e) Cross-sections of the diamond slab in the $x$-$z$ plane [black rectangle plane in (a)] with the electric field $\text{Re}(E_y(x,z))/\max(|E_y(x,z)|)$ overlaid. Slots are located at the even phase fronts (2, 4, ..., 26) in (b) for constructive interference. We add extra destructive interference slots around odd phase fronts in (c) (3), (d) (3 and 5), and (e) (3, 5, and 7). The black line shows the slot edges. (f) Slot locations and widths for the antenna designs in (b-e).}
\end{figure}

Step 1: calculate the field profile of $y$-oriented dipole, $ \text{Re}[E_y(x,y,0)]/\max(|E_y(x,y,0)|)$, in the 150 nm thick unpatterned diamond membrane, as shown in fig. 1(a). We define the $n^\text{th}$ phase front as the points with $n\pi$ phase difference from the dipole. 
The red arrows in fig. 1(a) indicate the even number of the phase fronts, which will provide constructive interference when adding dielectric perturbations to the diamond membrane. While the blue arrows indicate the odd number of the phase fronts, which will provide destructive interference with the scattering of the even number phase fronts when adding dielectric perturbations. We transform the target far field $\vec{e}_\text{tar}$ to the target near field $\vec{e}_\text{ntar}$, which has azimuthal symmetry in amplitude. We add dielectric perturbations along the phase fronts to make the mode normally incident on each perturbation layer. Each curved perturbation layer is then approximated by a straight slot or periodic array of holes. Under these approximations, the 3D design problem reduces to a 2D problem for slots ($x$-$z$ plane) or a 3D problem with a periodic boundary condition in the $y$ direction for the hole array.

Step 2: simulate an in-plane transverse-electric (TE) slab mode that is normally incident on a single slot (width $w$) or a single period of the hole array (diameter $d$, spacing $L$) with FDTD simulation. The simulation yields a lookup table with reflection and transmission coefficients as well as scattered near field distributions $\vec{E}_s(x)=\vec{E}_\text{near}(x,z_0)$ for slots or $\vec{E}_s(x) =\frac{1}{L}\int_0^L \vec{E}_\text{near}(x,y,z_0)\text{d}y$ for hole array spacing $L$. The $ i^\text{th} $ layer is at position $ x_i $ with slot width $ w_i $ or hole parameters $ (d_i, L_i) $. Here, $ i = 1, ..., N_\text{max} $, where $ N_\text{max} $ is the maximum number of the perturbation layers in the antenna. 

Step 3: apply transfer matrix model (TMM) to calculate the electric field at each layer~\cite{TMM,TMM2}. Thus, we obtain the total scattered near field by coherently adding contributions from each scattering layer using the lookup table. Here we use the slots located at even ($ 2, 4, 6,..., 26$) phase fronts of the dipole field in the unpatterned diamond slab as the initial guess structure.

Step 4: calculate the mode overlap between the target and the antenna-scattered near field along the line $ (x, y = 0, z = z_0) $. The parameters $ (x_i, w_i) $  or $(x_i, d_i, L_i) $ of each layer are optimized based on step 3 to maximize the mode overlap. 

Step 5: curve each slot or layer of holes to match the dipole emission phase fronts in the diamond membrane (centered at $ z = 0 $) and add a bottom reflector at $ z = -z_\text{min} $. We apply a gradient descent optimization to maximize $\eta$ calculated from 3D FDTD simulations using the result from step 4 as the initial guess structure. 

Step 6: add destructive interference slots (located at odd dipole field phase fronts like 3, 5, and 7) in the initial guess structure in step 3 and redo step 4 and step 5 to increase the antenna Purcell factor.

Figure 1(b-e) plots the optimized antenna $x$-$z$ cross-sections of step 4 overlaid with the electric field $\text{Re}(E_y(x,z))/\max(|E_y(x,z)|)$ without, with one, with two, and with three destructive interference slots in the initial guess structure, respectively. Figure 1(f) details the values for slot locations and widths in fig. 1(b-e). 

\section{Simulation Results}

\begin{figure}[htbp]
\centering
\includegraphics[width=13cm]{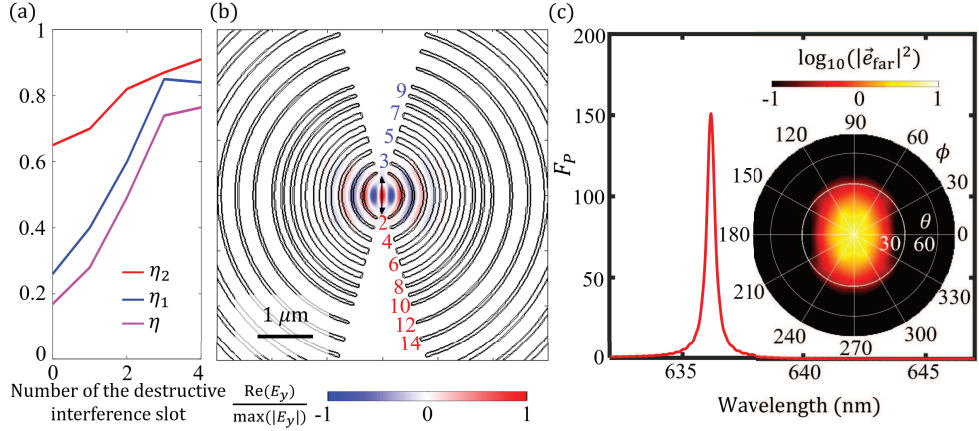}
\caption{ (a) Efficiency vs. number of the destructive interference slots. (b) $\text{Re}(E_y)/\max(|E_y|)$ of the antenna with four destructive interference slots (at phase fronts 3, 5, 7, and 9). The black line shows the edge of the slot in the antenna structure. The red texts label phase front number of the constructive interference slots while the blue texts label the destructive interference slots. (c) Purcell factor spectrum and the far-field distribution $\log_{10}(|\vec{e}_\text{far}|^2)$ of the antenna structure in (b).}
\end{figure}

Figure 2(a) plots the efficiency as a function of the number of the destructive interference slots. $\eta$ increases with the number of destructive interference slots but saturates after four slots since the field is confined in the central region as shown in fig. 2(b). Figure 2(b) shows the slot antenna design with an efficiency $\eta$ = 75\% for NVs using four destructive interference slots located at phase fronts (3, 5, 7, and 9) (See supplementary material for detailed geometries). Figure 2(c) plots the spectrum of the Purcell factor and far-field distribution of the antenna emission.


Next, we use arrays of holes as the perturbation layer instead of the slots in our recipe. The holes in the design have a minimum diameter ($70$ nm) than the minimum width of the slot ($40$ nm) relaxing fabrication difficulties. In addition, the design with holes ensures a connected suspended structure even if the layers wrap around $360^{\circ}$ with more degrees of freedom by hole radii and spacing. 

\begin{figure}[htbp!]
\centering
\includegraphics[width=13cm]{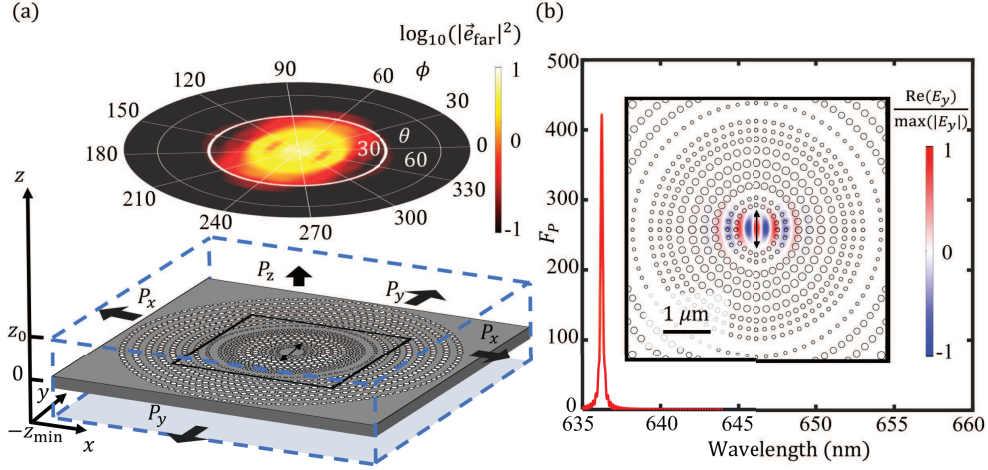}
\caption{ (a) Illustration of the dielectric antenna structure, along with a plot of $\log_{10}(|\vec{e}_\text{far}|^2)$ showing the far-field distribution. (b) Purcell factor spectrum of the antenna structure. The inset is a linear-scale plot of $\text{Re}(E_y)/\max(|E_y|)$ corresponding to the black square region in (a).}
\end{figure}

Figure 3(a) shows the holey dielectric antenna structure along with the emitted far field. Figure 3(b) plots the spectrum of the Purcell factor and electric field distribution in the antenna's $x$-$y$ cross-section. This structure simultaneously achieves a large mode overlap ($\eta_2$ reaches 87\%) and a large Purcell factor of 420, which makes the spin-photon interface $\eta$ reach 81\% for NVs. We optimized the structure for NV centers in diamond here. The design process can also work for other emitters by changing the target resonant wavelength. In the supplementary material, we also show an antenna design for the GaAs quantum dot system. In fig. 3(a), the distance between the emitter and the closest dielectric perturbation is larger than $\lambda/n_\text{d}$ (the resonant wavelength in the material, where $n_\text{d}$ is the refractive index of the diamond), which can alleviate surface-charge noise. A previously reported bullseye antenna design has a collection efficiency of 90\% within NA $= 0.65$ and a Purcell factor of 20 in a GaAs quantum dot system \cite{GaAs_Nat_Nano}. Our design applies the mode overlap between the antenna far field and the target Gaussian far field as the figure of merit to have a better estimation in single-mode fiber collection efficiency, which is important for quantum photonics, compared with the collection efficiency only considering the electromagnetic energy within a certain NA. Our design can simultaneously achieve a large mode overlap to a small NA mode and a large Purcell factor. The small NA collection can both provide lower magnification for a larger field of view to examine more quantum emitters and have a longer working distance between the objective lens and the cryogenic stage.

\section{Sensitivity Analysis}
The antenna design is tolerant to errors in dipole angle and location. Figure 4 plots the efficiencies and Purcell factor as a function of the dipole's misalignment in location $(\Delta x, \Delta y, \Delta z)$ and orientation, given by polar angle $ \theta $ and azimuth angle $\phi$. In fig. 3, the simulated dipole orientation is in the $y$ direction corresponding to $ \theta=0^\circ$ and $\phi=0^\circ$. The normalized Purcell Factor $f_p$ is 1 for the antenna structure shown in fig. 3. Figure 4(a) and 4(b) show how $\eta $, $ \eta_2 $, and $f_p$ vary with the dipole orientation. 

\begin{figure}[htbp]
\centering
\includegraphics[width=14cm]{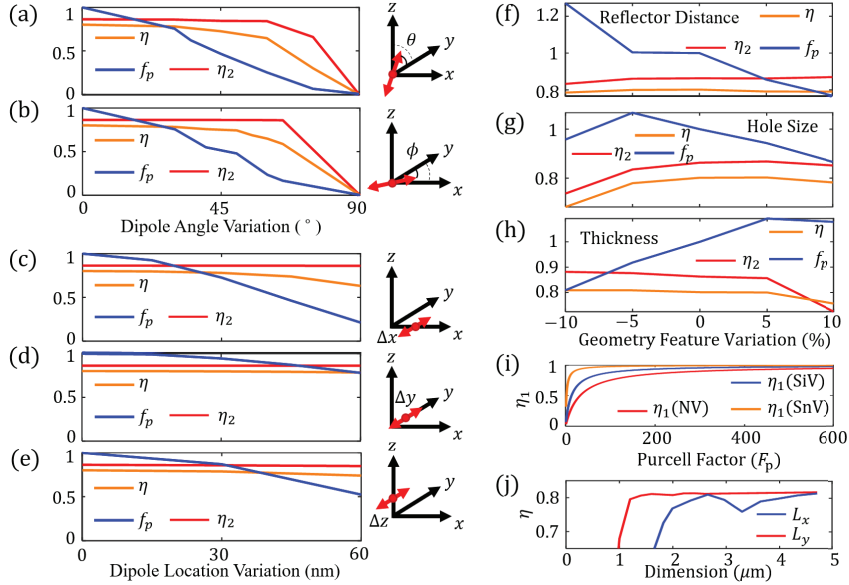}
\caption{The efficiencies $ \eta $ and $ \eta_2 $, as well as the normalized Purcell factor $f_p=F_p/\max(F_p)$ as a function of changes in the dipole angle $(\theta,\phi)$ (a, b), and dipole location  $(\Delta x,\Delta y,\Delta z)$ (c-e), bottom reflector location (f), hole size (g), and membrane thickness (h), respectively. (i) Dependence of $ \eta_1 $ on the Purcell factors for different types of quantum emitters. (j) Efficiency $\eta$ changes with $L_x$ and $L_y$.}
\end{figure}

Figure 4(c-e) summarize the effect of dipole displacements. The electric field changes rapidly in the $x$ direction while changing gradually in the $y$ direction, as seen in fig. 3(b). Figure 4(c,d) show that the Purcell factor decreases by 80\% for $\Delta x = 60$~nm, but only 20\% for $\Delta y = 60$~nm. 
The Purcell factor drops 50\% with $\Delta z = 60$ nm in fig. 4(e). For our antenna design for NVs, the target implantation depth is $75$~nm for the $150~$nm diamond slab, which corresponds to a 60 keV implantation energy in Stopping and Range of Ions in Matter (SRIM) simulation \cite{SRIM}. 
From the simulation, we are 95\% confident that the position variations of the implanted NV centers are within $|\Delta z|< 32$~nm and $\sqrt{(\Delta x^2+\Delta y^2)}<28$~nm under the Gaussian distribution assumption \cite{SRIM}. In summary, though the dipole location variation changes the Purcell factor, $\eta_2$ is changed less than 2\% when the angle variation is smaller than 45$^\circ$ since the dipole couples to the antenna mode and the antenna mode couples to the free space target mode. $f_p$ follows the expected overlap between the dipole $\vec{\mu}$ and the mode's electric field $\vec{E}$, i.e., $f_p \propto (\frac{|\vec{\mu} \cdot \vec{E}(r_i)|}{|\vec{\mu}|\cdot|\vec{E}_{max}|})^2$, where $\vec{E}(r_i)$ is the local electric field at the dipole emitter location $r_i$, and $|\vec{E}_{max}|$ is the maximum value of the electric field in the antenna \cite{faraon2011resonant}.

We also investigate variations in the bottom reflector distance, hole sizes, and membrane thickness (fig. 4(f-h)). We calculate $\eta$, $\eta_2$, and $f_p$ at the resonant wavelength in the simulation. The variation in the geometry will change the antenna resonant wavelength, but we can tune the resonant wavelength to the target value using e.g. gas tuning \cite{NV_Cavity}.  The antenna performance is robust to the bottom reflector distance as shown in fig. 4(f). From fig. 4(g) it is seen that increasing the hole size by 10\% only decreases the efficiency by 2\%, whereas a reduction in hole size by 10\% reduces the efficiency by 12\%. We observe a similar trend when decreasing rather than increasing the membrane thickness, as seen in fig. 4(h). 
The antenna efficiency ($\eta$) does not decrease more than 12\% for a $ \pm10\%$ geometry variation. The design with a large Purcell factor (fig. 4(i)) maintains $ \eta_1 $ even though the Purcell factor decreases by 70\% due to variations in dipole location, orientation, or nanostructure geometry. Within a $ \pm10\% $ geometry variation, $ \eta_2 $ does not decrease by more than 13\% when the dipole is coupled to the antenna mode, which leads to a robust spin-photon interface efficiency $ \eta $. 


Finally, we study the size dependence of the antenna. We reduce the antenna size to $2L_x~\times~2L_y$ where $L_x$ and $L_y$ are the length in the $x$ and $y$ direction. The simulation region is still $10~\mu$m$~\times~10~\mu$m, where the diamond slab dimension is $10~\mu$m$~\times~2L_y$. The regions $y>L_y$ and $y<-L_y$ are filled by air as the undercut trench \cite{wan2018two}. In fig. 4(j), $\eta$ will not reduce more than 5\% when $L_x>2~\mu$m fixing $L_y=5~\mu$m. When reducing $L_y$ with the fixed $L_x=5~\mu$m, we notice that $\eta$ does not decrease more than 2\% when $L_y>1.2~\mu$m. The $\eta$ is over 73\% with dimension $4~\mu$m$~\times~2.4~\mu$m for an NV center which resonant wavelength is $\lambda=637$ nm.

\section{Conclusion}
In conclusion, we introduced a transfer-matrix approach for cavity-grating designs, enabling efficient calculation of the cavity mode and the far field after an FDTD simulation of scattering matrix primitives. We applied this method to design an efficient dielectric antenna for quantum emitters. Specifically, the design achieves: (i) a relatively large emitter spacing to the first etched surface of 1.4 $\lambda/n_\text{d}$ to alleviate surface charge noise; (ii) 93\% mode overlap with a 0.4 NA Gaussian beam which has 99\% collection efficiency within an NA of 0.5, together with a Purcell factor of 420; (iii) robustness to fabrication and dipole variations. While we considered a diamond membrane here, the design applies to diverse materials such as Si or GaAs. We anticipate that this design methodology and the resulting efficient quantum emitter interfaces will benefit numerous applications, including multiplexed quantum repeaters\cite{chen}, arrayed quantum sensors\cite{sensing,zagoskin2013spatially}, boson sampling ~\cite{Science_boson_2020}, and spin-based fault-tolerant quantum computers \cite{choi19}.

\section*{Acknowledgments}
The authors thank H. Raniwala and M. ElKabbash for useful comments. The authors acknowledge the support from the Defense Advanced Research Projects Agency (DARPA) DRINQS (HR001118S0024), the Air Force Office of Scientific Research MURI (FA9550-14-1-0052), and the MITRE Quantum Moonshot Program. L. L. acknowledges support from the Analog Devices Fellowship, EFMA-1641064, QISE-NET NSF award DMR-1747426, and the NSF CIQM program. H. C. acknowledges the Claude E. Shannon Fellowship and Samsung Scholarship.



\bibliography{antenna}



\end{document}